\documentclass[11pt]{article}
\usepackage{calc}
\usepackage{xspace}
\usepackage{amsmath}
\usepackage{amssymb}
\usepackage{amsfonts}
\usepackage{mathrsfs}
\usepackage{enumerate}
\makeatletter
\@addtoreset{equation}{section}
\makeatother
\usepackage{feynmf}
\usepackage{graphicx}
\usepackage{amssymb,amsfonts,amsmath}
\usepackage{accents}

\def\beq{\begin{equation}}
\def\eeq{\end{equation}}
\def\bea{\begin{eqnarray}}
\def\eea{\end{eqnarray}}
\def\Tr{{\rm Tr}}

\allowdisplaybreaks

\begin{document}

\renewcommand{\thefootnote}{\fnsymbol{footnote}}
\newcommand{\inst}[1]{\mbox{$^{\text{\textnormal{#1}}}$}}
\begin{flushright}
EPHOU 16-014\\
Sept. 2016
\end{flushright}
\mbox{}\\\bigskip\bigskip
\begin{center}
{\LARGE  Quaternion based generalization of Chern-Simons theories in arbitrary
dimensions }\\[8ex]
%
{\large
Alessandro D'Adda\inst{a}\footnote{\texttt{dadda@to.infn.it}},
Noboru Kawamoto\inst{b}\footnote{\texttt{kawamoto@particle.sci.hokudai.ac.jp}},
Naoki Shimode\inst{b}\footnote{\texttt{nshimode@particle.sci.hokudai.ac.jp}}, 
\\[1ex] 
{\normalsize and}
Takuya Tsukioka\inst{c}\footnote{\texttt{tsukioka@bukkyo-u.ac.jp}}}\\[4ex]
%
{\large\itshape
\inst{a} INFN Sezione di Torino, and\\
Dipartimento di Fisica Teorica,
Universita di Torino, \\
I-10125 Torino, Italy\\[3ex]
\inst{b} Department of Physics, Hokkaido University, \\
Sapporo 060-0810, Japan\\[3ex]
\inst{c} School of Education, Bukkyo University,\\ 
Kyoto 603-8301, Japan}
\end{center}
\bigskip\bigskip
\setcounter{footnote}{0}
\renewcommand{\thefootnote}{\arabic{footnote}}

\begin{abstract}
A generalization of Chern-Simons gauge theory is formulated in any 
dimension and arbitrary gauge group where gauge fields and gauge 
parameters are differential forms of any degree.
The quaternion algebra structure of this formulation is shown 
to be equivalent to a three $\mathbb{Z}_2$-gradings structure, thus clarifying 
the quaternion role in the previous formulation. 
\end{abstract}




\newpage

\noindent
{\bf 1 \ Introduction}

\vspace*{1mm}

\noindent
A formulation of gauge theory in terms of differential forms has the advantage 
that it automatically generates a general coordinate invariant formulation 
since an explicit metric dependence does not appear. One interesting 
approach along this line is the applications of Chern-Simons action 
to 3-dimensional gravity~\cite{Witten-3dgravity}. 

In the formulation of the standard 
gauge theory only 1-form gauge fields and 0-form gauge parameters play 
a role as differential forms. It is natural to ask if one can formulate 
gauge theories in terms of all the degrees of differential forms. 
A positive answer was given by one of authors (N.K.) and Watabiki many years 
back with a graded Lie algebra setting~\cite{kawa-wata}. 
In this paper we focus on the generalization of the Chern-Simons 
action to arbitrary dimensions 
with arbitrary degrees of differential forms as gauge fields and parameters 
for Lie algebra setting and clarify the origin of the quaternion 
structure which was discovered in the original 
formulation~\cite{kawa-wata}. 
In the present formulation the introduction of graded Lie algebra is not 
required. 

\vspace*{3mm}

\noindent
{\bf 2 \ The origin of the quaternion structure for a three grading
gauge system}

\vspace*{1mm}

\noindent
When we consider standard Abelian gauge theory with differential forms 
we identify gauge 
field as one-form and gauge parameter as zero-form. In this gauge system 
$\mathbb{Z}_2$-grading structure of even-form and odd-form is present. If we define 
$\Lambda_+$ as a set of even forms and $\Lambda_-$ as a set of odd forms, 
we have
\beq
\lambda_+ \wedge \lambda'_+=\lambda'_+ \wedge \lambda_+\in\Lambda_+, 
\quad
\lambda_+ \wedge \lambda_-=\lambda_- \wedge \lambda_+\in\Lambda_-,
\quad
\lambda_- \wedge \lambda'_-=-\lambda'_- \wedge \lambda_-\in\Lambda_+, 
\label{z2grading}
\eeq
where $\lambda_+, \lambda'_+ \in \Lambda_+,~ \lambda_-, \lambda'_- \in 
\Lambda_-$ and $\wedge$ is a wedge product. Fermionic and bosonic 
fields have similar $\mathbb{Z}_2$-grading structure with an obvious 
correspondence.

Let us consider two types of fields $\Phi_{(a,b,c)}$ 
and ${\mathcal F}_{(a,b,c)}$ 
which have a three $\mathbb{Z}_2$-grading structure $(a,b,c)$ with $a,b,c, =0$ or 1. 
For simplicity we assume that these fields have Abelian nature. 
Then we introduce two types of commuting structure with respect to the three 
gradings:
\bea
\Phi_{(a,b,c)} \Phi'_{(a',b',c')}\!\!\!\! &=& \!\!\!\! 
(-1)^{aa' +bb'+cc'} \Phi'_{(a',b',c')} \Phi_{(a,b,c)}, 
\label{grading1}\\
{\mathcal F}_{(a,b,c)}  {\mathcal F}'_{(a',b',c')} \!\!\!\! &=& \!\!\!\!
(-1)^{(a+b+c)(a'+b'+c')} {\mathcal F}'_{(a',b',c')} {\mathcal F}_{(a,b,c)}.
\label{grading2}
\eea
In  (\ref{grading1}) the three gradings are independent whereas 
in (\ref{grading2}) there is one global grading corresponding 
$a+b+c$. 
Let us introduce an object ${\bf q}(a,b,c)$ satisfying 
the following commuting structure:
\beq
{\bf q}(a,b,c) {\bf q}(a',b',c') 
= (-1)^{aa'+bb'+cc' +(a+b+c)(a'+b'+c')} {\bf q}(a',b',c') {\bf q}(a,b,c). 
\label{quaternion-grading}
\eeq
Then we have the following commuting relation: 
\bea
(\Phi_{(a,b,c)}{\bf q}(a,b,c))\!\!\!\!\!\!\!\!&&
\!\!\!\!\!\!\!\!(\Phi'_{(a',b',c')}{\bf q}(a',b',c')) 
 \nonumber \\
&=&\!\!\!\!(-1)^{(a+b+c)(a'+b'+c')} \big(\Phi'_{(a',b',c')}{\bf q}(a',b',c')\big)
 \big(\Phi_{(a,b,c)} {\bf q}(a,b,c)\big),
 \label{product-grading}
\eea
where we assume $\Phi_{(a,b,c)}$ and ${\bf q}(a,b,c)$ are not interacting 
and thus commuting:
\beq
\Phi_{(a,b,c)} {\bf q}(a,b,c) = {\bf q}(a,b,c) \Phi_{(a,b,c)}. 
\label{q-commuting}
\eeq
We recognize now that a field of the ${\mathcal F}$-type, namely 
with a single global grading, can be written in terms of a field 
$\Phi_{(a,b,c)}$ with three distinct gradings as
\beq
{\mathcal F}_{(a,b,c)} =\Phi_{(a,b,c)} {\bf q}(a,b,c). 
\label{q-valued-field} 
\eeq
In order to define a product of the ${\mathcal F}$-fields, we have 
to define a product in the ${\bf q}(a,b,c)$ space that satisfies 
(\ref{quaternion-grading}). This is given by
\beq
{\bf q}(a,b,c){\bf q}(a',b',c') = (-1)^{aa'+bb'+cc'+a'b+b'c+c'a} 
{\bf q}(a+a',b+b',c+c'), 
\label{quaternion-closure}
\eeq
where the sums are defined modulo 2. This product is associative. 
There are eight possible 
${\bf q}(a,b,c)$'s for $a,b,c = 0,1$. However with respect to their 
commuting structure (\ref{quaternion-grading}) and with respect to the 
product (\ref{quaternion-closure}) there are only four independent 
${\bf q}(a,b,c)$. In fact the sign factors in (\ref{quaternion-grading}) 
and (\ref{quaternion-closure}) are invariant for 
$(a,b,c) \rightarrow (a+1,b+1,c+1)$ and the same for $(a',b',c')$. 
The units ${\bf q}(a,b,c)$ can then be identified in pairs and renamed as:
\begin{equation}
\renewcommand{\arraystretch}{1.4}
\begin{array}{rcl}
{\bf q}(1,1,1) = {\bf q}(0,0,0) \equiv {\bf 1}, &\quad&
{\bf q}(1,0,0) = {\bf q}(0,1,1) \equiv {\bf i}, \\
{\bf q}(0,1,0) = {\bf q}(1,0,1) \equiv {\bf j}, &\quad&
{\bf q}(0,0,1) = {\bf q}(1,1,0) \equiv {\bf k}.
\end{array}
\renewcommand{\arraystretch}{1}
\label{quat-def}
\end{equation}
It is easy to recognize now that ${\bf 1,i,j,k}$ satisfy the quaternion 
algebra 
\begin{equation}
\renewcommand{\arraystretch}{1.4}
\begin{array}{c}
{\bf i}^2 = {\bf j}^2 = {\bf k}^2 = -{\bf 1},
 \\ 
{\bf i}{\bf j} = -{\bf j}{\bf i} = {\bf k}, \ \ 
{\bf j}{\bf k} = -{\bf k}{\bf j} = {\bf i}, \ \ 
{\bf k}{\bf i} = -{\bf i}{\bf k} = {\bf j}.
\end{array} 
\renewcommand{\arraystretch}{1}
\label{quat-com-rel}
\end{equation}
It is important to notice that eq.(\ref{quaternion-closure}) defines the 
quaternion algebra in an 
unconventional way, different from its standard mathematical introduction, and 
that it links in an unexpected way the quaternion algebra to the existence of 
three independent gradings.

We can introduce now two types of ${\mathcal F}$-fields, corresponding 
respectively to $a+b+c$ odd and even:
\bea
{\mathcal A} \!\!\!\!&=&\!\!\!\!{\bf q}(1,1 ,1) {\Phi}_{(1,1,1)}  +
{\bf q}(1,0,0) {\Phi}_{(1,0,0)}
  +  {\bf q}(0,1,0) {\Phi}_{(0,1,0)}  +
{\bf q}(0,0,1) {\Phi}_{(0,0,1)} \nonumber \\
\!\!\!\!&=&\!\!\!\!{\bf 1} {\Phi}_{(1,1,1)}  +   {\bf i} {\Phi}_{(1,0,0)}
  +  {\bf j} {\Phi}_{(0,1,0)}  +  {\bf k} {\Phi}_{(0,0,1)},
\label{gene-field}\\
{\mathcal V}
\!\!\!\!&=&\!\!\!\!
{\bf q}(0,0,0) {\Phi}_{(0,0,0)}  +   {\bf q}(0,1,1) {\Phi}_{(0,1,1)}
  +  {\bf q}(1,0,1) {\Phi}_{(1,0,1)}  +  {\bf q}(1,1,0) {\Phi}_{(1,1,0)} \nonumber \\
\!\!\!\!&=&\!\!\!\!
{\bf 1} {\Phi}_{(0,0,0)}  +   {\bf i} {\Phi}_{(0,1,1)}
  +  {\bf j}{\Phi}_{(1,0,1)}  +  {\bf k} {\Phi}_{(1,1,0)}.
 \label{gene-para}
\eea
Assuming that the fields $\Phi_{(a,b,c)}$ are Abelian ${\mathcal A}$ 
and ${\mathcal V}$ are odd and even elements in a $\mathbb{Z}_2$ 
commuting algebra: 
\beq
{\mathcal A} {\mathcal A}'=-{\mathcal A}'{\mathcal A}, \ \ \ \
{\mathcal A}{\mathcal V}  = {\mathcal V}{\mathcal A}, \ \ \ \
{\mathcal V}{\mathcal V}' =  {\mathcal V}'{\mathcal V},
\label{gene-z2commuting} 
\eeq
where ${\mathcal A}'$ and ${\mathcal V}'$ are defined by $\Phi'(a',b',c')$. 

It also follows immediately from (\ref{gene-field}), (\ref{gene-para}) 
and the multiplication rules of quaternions that if we denote by 
${\bf\Lambda}_-$ and ${\bf\Lambda}_+$ the sets of fields respectively of the 
${\mathcal A}$-type and ${\mathcal V}$-type then 
\bea
{\mathcal A} {\mathcal A}'=-{\mathcal A}'{\mathcal A} \in {\bf\Lambda}_+, \ \ \
{\mathcal A}{\mathcal V}  = {\mathcal V}{\mathcal A} \in {\bf\Lambda}_-,  \ \ \
{\mathcal V}{\mathcal V}' =  {\mathcal V}'{\mathcal V} \in {\bf\Lambda}_+. 
\label{gene-z2grading}
\eea

In conclusion ${\mathcal A}$ and ${\mathcal V}$ are anticommuting 
and commuting quaternionic fields whose component fields 
$\Phi_{(a,b,c)}$ possess three independent $\mathbb{Z}_2$-gradings whose 
physical meaning and interpretation may be different. 
We shall use them in what follows to formulate gauge theories 
with higher differential forms. 

The importance of the different sign choice in 
(\ref{grading1}) and (\ref{grading2}) was noticed in 
\cite{Deligne} for two- and three-gradings formulations 
of supersymmetric gauge theories. 

\vspace*{3mm}

\noindent
{\bf 3 \ Higher form gauge fields and non-Abelian extension}

\vspace*{1mm}

\noindent
We shall consider fields $\Phi_{(a,b,c)}(x)$ with space-time 
dependence so it is natural to associate one of the 
gradings, -conventionally the second one labeled by $b$-, as 
representing the grading of even and odd differential forms in 
space-time. We shall also take the first grading, labeled by 
$a$, as some sort of fermion-boson grading (whose exact nature 
will be briefly discussed at the end) while the third grading 
will be left unspecified and denoted by the suffix $c(=0,1)$.
 
We introduce then the following notations:
\begin{equation}
\renewcommand{\arraystretch}{1.4}
\begin{array}{rcl}
\Phi_{(0,0,c)}(x) \!\!\!\!&\equiv&\!\!\!\! \hat{A}_c(x)={\hbox{direct sum of
bosonic even forms}}, \\
\Phi_{(0,1,c)}(x) \!\!\!\!&\equiv&\!\!\!\! A_c(x) ={\hbox{direct sum of bosonic odd
 forms}}, 
\\
\Phi_{(1,0,c)}(x) \!\!\!\!&\equiv&\!\!\!\! \hat{\psi}_c(x)={\hbox{direct sum of
fermionic even forms}},  \\
\Phi_{(1,1,c)}(x) \!\!\!\!&\equiv&\!\!\!\! \psi_c(x) = {\hbox{direct sum of fermionic 
odd forms}}. 
\end{array}
\renewcommand{\arraystretch}{1}
\label{grading-notation}
\end{equation}
Here and in the following we shall denote bosons and fermions, 
respectively, with Roman and Greek letters while even forms and odd 
forms are differentiated by hat and non-hat.  

We introduce the exterior derivative operator $d=dx^\mu\partial_\mu$ 
as an ${\mathcal A}$-type operator since it has grading; $(0,1,0)$: 
\beq
Q= {\bf q}(0,1,0) d ={\bf j} d.  
\label{gene-d-op}
\eeq
A more general form of $Q$ will be briefly discussed at the end of 
the paper. 

Extending from Abelian to non-Abelian gauge field theory, we identify 
the following ${\mathcal A}$ and ${\mathcal V}$ as generalized gauge field 
and gauge parameter respectively:
\begin{equation}
\renewcommand{\arraystretch}{1.4}
\begin{array}{rcl}
{\mathcal A} \!\!\!\!&=&\!\!\!\! {\mathcal A}^B T^B = ({\bf 1} {\psi}_1^B  +
{\bf i} \hat{\psi}_0^B   +  {\bf j} {A}_0^B  +  {\bf k} \hat{A}_1^B)T^B,  
\\
{\mathcal V}\!\!\!\!&=&\!\!\!\! {\mathcal V}^B T^B = ({\bf 1} \hat{a}_0^B  +   
{\bf i} {a}_1^B   +  {\bf j} \hat{\alpha}_1^B  +  {\bf k} {\alpha}_0^B) T^B,
\end{array}
\renewcommand{\arraystretch}{1}
\label{gene-nonAbelian}
\end{equation}
where $T^B$'s are generators of Lie algebra. It is important to realize at this stage that 
${\mathcal A}^B$ and ${\mathcal V}^B$ satisfy the same grading structure 
as ${\mathcal A}$-type and ${\mathcal V}$-type in (\ref{gene-z2grading}). 
Hereafter all products such as 
${\mathcal A}^A{{\mathcal A}'}^B,~{\mathcal A}^A{\mathcal V}^B,~
{{\mathcal V}'}^A{\mathcal V}^B \cdots$ are 
understood as wedge products. 

It is important to remark at this stage that ${\mathcal A}$, ${\mathcal V}$ 
and $Q$ have the same formal properties as respectively one form gauge 
fields, zero form gauge parameters and the differential operator $d$ in 
gauge theories. So any gauge theory that can be entirely written 
in terms of forms, without use of the Hodge operator, 
like Chern-Simons theory and Einstein gravity, admits a generalization 
where all fields and parameters are replaced by the quaternionic analogues 
${\mathcal A}$ and ${\mathcal V}$ and gauge invariance in terms of the 
generalized parameters is automatically preserved. 
In this paper we shall concentrate on the generalized Chern-Simons 
theory, which was already studied in a similar framework in \cite{kawa-wata}.

\vspace*{3mm}

\noindent
{\bf 4 \ Generalized Chern-Simons actions in arbitrary dimensions}

\vspace*{1mm}

\noindent
We can now construct a dimension independent formulation of generalized 
Chern-Simons action: 
\beq
S = \int \Tr\left(\frac{1}{2}{\mathcal A}Q{\mathcal A} + 
\frac{1}{3}{\mathcal A}^3  \right) =
{\bf 1}S^1 + {\bf i}S^i + {\bf j}S^j + {\bf k}S^k,   
\label{GCSA}
\eeq
where the $\Tr$ is taken on a representation of the Lie algebra. 
Notice that ${\mathcal A}$ contains now forms of arbitrary degree and 
so the integral is not restricted to be on a three dimensional 
space-time but can be formulated in any dimensions. 
This generalized Chern-Simons action is invariant under the following 
generalized gauge transformation:
\beq
\delta {\mathcal A} = Q{\mathcal V} +[{\mathcal A},{\mathcal V}] 
={\bf 1}\delta \psi_1 + {\bf i}\delta \hat{\psi}_0 + 
{\bf j}\delta A_0 + {\bf k}\delta\hat{A}_1.
\label{GGT}
\eeq

The proof of the gauge invariance of the generalized Chern-Simons action 
can be derived from the following properties of the generalized 
gauge fields and parameters: 
\begin{enumerate}
\item $Q^2 =0$,
\item $\{\overrightarrow{Q},{\bf \lambda_-} \} = Q{\bf \lambda_-}, ~~~
[\overrightarrow{Q},{\bf \lambda_+}] = Q{\bf \lambda_+},$ 
\item $\Tr({\bf \lambda_+}{\bf \lambda'_+}) =
\Tr({\bf \lambda'_+}{\bf \lambda_+}), ~~~
\Tr({\bf \lambda_-}{\bf \lambda_+}) =
\Tr({\bf \lambda_+}{\bf \lambda_-}), ~~~
\Tr({\bf \lambda_-}{\bf \lambda'_-}) =-
\Tr({\bf \lambda'_-}{\bf \lambda_-})$,
\end{enumerate}
where $\lambda_-,\lambda'_-$ and $\lambda_+,\lambda'_+$ are, 
respectively, ${\mathcal A}$-type 
and ${\mathcal V}$-type fields and parameters in (\ref{gene-nonAbelian}).
Here $\{\ ,\ \}$ and $[\ , \ ]$ are anti-commutator and commutator, respectively. 
$\overrightarrow{Q}={\bf j}\overrightarrow{d}$ is an exterior derivative 
operating on the right. 

In this quaternion valued formulation each sector of quaternion coefficients 
are equivalent in (\ref{GCSA}) and (\ref{GGT}) since quaternions commute 
with fields and interact only among themselves. 
We can thus derive the following four types of generalized Chern-Simons 
actions:
\bea
S^1\!\!\!\!&=&\!\!\!\! 
\int \Tr\left[ -\psi_1(dA_0+A_0^2+\hat{A}_1^2+\hat{\psi}_0^2)
+\hat{A}_1(d \hat{\psi}_0+[A_0,\hat{\psi}_0]) + \frac{1}{3}\psi_1^3 \right],
\nonumber \\
S^i \!\!\!\!&=&\!\!\!\!
\int\Tr\left[-\hat{\psi}_0(dA_0+A_0^2+\hat{A}_1^2- \psi_1^2)  
 -\hat{A}_1(d\psi_1+\{A_0,\psi_1\}) - \frac{1}{3}\hat{\psi}_0^3    \right] ,
 \nonumber \\
 S^j \!\!\!\!&=&\!\!\!\!
\int \Tr\left[ -\frac{1}{2}A_0 d A_0 -\frac{1}{3} A_0^3
+ \frac{1}{2}\hat{A}_1(d \hat{A}_1 +[A_0 , \hat{A}_1])
 \right.
\label{GCSAs}
\\
 &&
\hspace*{8mm}
\left.
   +\frac{1}{2}\hat{\psi}_0(d \hat{\psi}_0 +[A_0, \hat{\psi}_0]) 
  + \frac{1}{2} \psi_1(d\psi_1+\{ A_0,\psi_1\})
 - \hat{\psi}_0\{\psi_1,\hat{A}_1\}
  \right] ,
 \nonumber \\
 S^k \!\!\!\!&=&\!\!\!\!
\int \Tr\left[- \hat{A}_1(dA_0+A_0^2 +\hat{\psi}_0^2-  \psi_1^2) 
 -\frac{1}{3} \hat{A}_1^3 -\psi_1(d \hat{\psi}_0 +[A_0,\hat{\psi}_0]) \right] ,
\nonumber 
\eea
which are invariant under the following generalized gauge transformations:
\begin{equation}
\renewcommand{\arraystretch}{1.4}
\begin{array}{rcl}
\delta A_0 \!\!\!\!&=&\!\!\!\!d \hat{a}_0 +[A_0, \hat{a}_0] +\{\hat{A}_1,a_1\}
+[\psi_1,\hat{\alpha}_1] -\{\hat{\psi}_0,\alpha_0\}  ,
\\
\delta \hat{A}_1\!\!\!\!&=&\!\!\!\! -d a_1 -\{A_0,a_1\} + [\hat{A}_1, \hat{a}_0]
+ [\psi_1,\alpha_0] +\{\hat{\psi}_0,\hat{\alpha}_1\} ,
\\
\delta\psi_1\!\!\!\!&=&\!\!\!\!
-d \hat{\alpha}_1 -[A_0,\hat{\alpha}_1] -[\hat{A}_1,\alpha_0]
+[\psi_1, \hat{a}_0] - [\hat{\psi}_0, a_1] , 
\\
\delta\hat{\psi}_0 \!\!\!\!&=&\!\!\!\!
d \alpha_0 +\{A_0,\alpha_0\} -\{\hat{A}_1,\hat{\alpha}_1\}
+[\psi_1,a_1] + [\hat{\psi}_0,\hat{a}_0].
\end{array}
\label{GGTs}
\renewcommand{\arraystretch}{1}
\end{equation}
In these generalized gauge transformations we find that commutators 
and anti-commutators are mixed. It is, however, important to realize that the 
generators of the Lie algebra appear only in commutators and thus are 
algebraically closed. 
For example $\{A_0,\alpha_0\} =A_0^B \alpha_0^C[T^B,T^C]$, \
$\{\hat{\psi}_0,\hat{\alpha}_1\} = \hat{\psi}_0^B\hat{\alpha}_1^C[T^B,T^C]$, \ 
$\{\hat{A}_1,a_1\} = \hat{A}_1^B a_1^C[T^B,T^C], \ \cdots$ and so on. 
Notice that the standard 3-form Chern-Simons action and the standard 
gauge transformation are included respectively in the first two terms of 
$S^j$ and of $\delta A_0$. 

It is important to recognize that the integrand of the generalized 
Chern-Simons action has the ${\mathcal A}$-type nature given 
in (\ref{gene-field}) with respect to the quaternion structure. 
$S^j$ is thus bosonic odd-dimensional action and $S^k$ is bosonic 
even-dimensional action. $S^1$ and $S^i$ are, respectively, fermionic 
odd-dimensional and fermionic even-dimensional actions, whose physical 
interpretation is not yet clear at this moment. 
Notice that the structure of $S^j$ is different from the one of the 
other actions $S^1,S^i$ and $S^k$ which have instead a similar 
field structure. The origin of this difference and 
similarity comes respectively from the special choice of the exterior 
derivative operator as a ${\bf j}$-component quaternion; $Q={\bf j}d$ 
and from the permutation invariance 
in the quaternion space. 

\vspace*{3mm}

\noindent
{\bf 5 \ Connection with graded Lie algebra formulation}

\vspace*{1mm}

\noindent
The quaternion structure in formulating higher form 
gauge theory was discovered long time ago by one of 
authors (N.K.) and Watabiki~\cite{kawa-wata}. 
In fact the generalized 
Chern-Simons actions (\ref{GCSAs}) and the gauge transformations 
(\ref{GGTs}) were already given in \cite{kawa-wata} in a  
graded Lie algebra framework. At the time the origin of quaternions in 
formulating the generalized gauge theory was not clear. 
In the present paper the origin of the quaternion is clarified and is based 
on the 3-grading structure. Due to this clarification a Lie algebra 
formulation instead of graded Lie algebra formulation has been 
successfully realized. 
Here we show how these two formulations are related in the current context. 

In order that a product of algebra valued fields and parameters of 
${\mathcal A}$-type and ${\mathcal V}$-type again belongs to 
${\mathcal A}$-type or ${\mathcal V}$-type field or parameter, 
the products should be defined by the following graded commutator:
\beq
{\mathcal A}\cdot {\mathcal A}'\equiv 
\frac{1}{2}\{{\mathcal A},{\mathcal A}' \},\quad 
{\mathcal A}\cdot {\mathcal V}\equiv 
\frac{1}{2}[{\mathcal A},{\mathcal V} ],\quad 
{\mathcal V}\cdot {\mathcal V}' \equiv 
\frac{1}{2}[{\mathcal V},{\mathcal V}' ],
\label{closure-alg}
\eeq
where ${\mathcal A}, {\mathcal A}'$ and ${\mathcal V}, {\mathcal V}'$ are 
(graded) Lie algebra valued ${\mathcal A}$- and ${\mathcal V}$-type fields 
and/or parameters, respectively. In \cite{kawa-wata} 
the necessity of algebraic closure for the product was noticed but 
not clearly stated while this point was stressed in \cite{Schwarz}.

Let us now consider the case where we do not introduce the third grading 
denoted by the suffix $c=(0,1)$, but introduce two types of generators 
$T^B$ and $\Sigma^\alpha$ which close as a graded Lie algebra:
\beq
[T^B,T^C]= f^{BC}_D T^D, ~~~
[T^B,\Sigma^\alpha] = g^{B\alpha}_\beta \Sigma^\beta,~~~
\{\Sigma^\alpha,\Sigma^\beta\} = h^{\alpha \beta}_B T^B,
\label{GLA}
\eeq
where $f^{BC}_D,g^{B\alpha}_\beta$ and $h^{\alpha \beta}_B$ are the 
structure constants of the graded Lie algebra. 

If we define ${\mathcal A}$-type and ${\mathcal V}$-type graded Lie algebra 
valued fields and/or parameters as:
\begin{equation}
\renewcommand{\arraystretch}{1.4}
\begin{array}{rcl}
{\mathcal A}\!\!\!\!&=&\!\!\!\!
{\bf 1}\psi^\alpha\Sigma^\alpha + {\bf i}\hat{\psi}^BT^B
+{\bf j}A^BT^B +{\bf k}\hat{A}^\alpha\Sigma^\alpha
\in {\bf \Lambda}'_-, \\
{\mathcal V}\!\!\!\!&=&\!\!\!\!
{\bf 1}\hat{a}^BT^B  + {\bf i}a^\alpha\Sigma^\alpha
+{\bf j}\hat{\alpha}^\alpha\Sigma^\alpha +{\bf k}\alpha^BT^B
\in {\bf \Lambda}'_+,
\end{array}
\label{defAV-GLA}
\end{equation}
where ${\bf \Lambda}'_-$ and ${\bf \Lambda}'_+$ denote the set of 
${\mathcal A}$-type and ${\mathcal V}$-type fields and/or parameters, 
respectively. 
We can then show
\beq
{\mathcal A}\cdot {\mathcal A}' \in {\bf \Lambda}'_+, \quad 
{\mathcal A}\cdot {\mathcal V} \in {\bf \Lambda}'_-, \quad 
{\mathcal V}\cdot {\mathcal V}'\in {\bf \Lambda}'_+,
\label{closure-prod-GLA}
\eeq
which are the extended version of the closure relation (\ref{gene-z2grading}) 
for graded Lie algebra.  

In order to obtain the gauge invariance of the generalized Chern-Simons action 
(\ref{GCSA}) for the graded Lie algebra framework, it is necessary the item 
3. in section 4 to be fulfilled for graded Lie algebra counterparts: 
\beq
 {\rm Str}[{\mathcal V},{\mathcal V}'] = {\rm Str}
[{\mathcal A},{\mathcal V}] =
{\rm Str}\{{\mathcal A},{\mathcal A}'\} =0.
\label{Str-def}
\eeq
All these conditions can be satisfied if super trace (Str) satisfies the 
following relations for graded generators: 
\beq
{\rm Str}[T^B,T^C] = {\rm Str}[T^B,\Sigma^\alpha] =
{\rm Str}\{\Sigma^\alpha,\Sigma^\beta\}=0.
\label{Str-relations}
\eeq
In particular the odd generators $\Sigma^\alpha$'s anti-commute 
within the super trace. 

This result shows that the 3rd grading with Lie algebra formulation is 
equivalent to the graded Lie algebra formulation without 3rd grading but 
with the super trace relations (\ref{Str-relations}). 

In the earlier generalized gauge theory formulation, the introduction of 
graded Lie algebra is necessary to formulate odd dimensional generalized 
Chern-Simons actions including all degrees of differential forms equally 
successful as for the even dimensional case~\cite{kawa-wata}. 
In other words the 3rd grading of the current formulation is mandatory 
for a equal footing treatment of generalized Chern-Simons actions 
for both even and odd dimensions. Thus introduction of three grading, 
equivalently the quaternion structure, is very important for the 
dimension independent treatment of the generalized gauge theory.  

As an application of the above graded Lie algebra formulation, 
Clifford algebra formulation which satisfies loosened version of 
the condition (\ref{closure-alg}); a product of $T^B$ and 
$\Sigma^\alpha$ generators close within product, 
was investigated to formulate conformal gravity in two and 
four dimensions~\cite{kawa-wata2}.


\vspace*{3mm}

\noindent
{\bf 6 \ D-dimensional generalized Chern-Simons actions}

\vspace*{1mm}

\noindent
It is important to realize that in  the generalized Chern-Simons action 
the generalized gauge invariance is valid order by order in the form 
degree. In other words 0-form, 1-form, 2-form, 3-form, 
4-form,$\cdots$ sectors of generalized Chern-Simons actions are separately 
invariant under the generalized gauge transformations of the 
gauge fields. 

To derive explicit forms of the generalized Chern-Simons actions 
with third grading $(c=0,1)$ for Lie algebra setting we 
introduce the following notations to clarify the differential form 
degrees: 
\begin{equation}
\renewcommand{\arraystretch}{1.4}
\begin{array}{rcl}
{\mathcal A} \!\!\!\!&=&\!\!\!\! {\bf 1} {\psi}_1  +  {\bf i} \hat{\psi}_0  
 +  {\bf j} {A}_0  +  {\bf k} \hat{A}_1
\\
&=&\!\!\!\!{\bf 1}(\psi_1^{(1)} + \psi_1^{(3)} +\cdots )
     + {\bf i} (\psi_0^{(0)} + \psi_0^{(2)} +\psi_0^{(4)} +\cdots )
\\
 & &\!\!\!\! +{\bf j}(\omega_0^{(1)} + \Omega_0^{(3)} + \cdots )
     +{\bf k} (\phi_1^{(0)} + B_1^{(2)} + H_1^{(4)}+ \cdots ),
\\
{\mathcal V}\!\!\!\! &=&\!\!\!\!  {\bf 1} \hat{a}_0  + {\bf i} {a}_1
+  {\bf j} \hat{\alpha}_1  +  {\bf k} {\alpha}_0 
\\
&=&\!\!\!\! {\bf 1}(v_0^{(0)} + b_0^{(2)} + h_0^{(4)} +\cdots ) 
  +{\bf i}(u_1^{(1)} + U_1^{(3)} +\cdots ) 
\\
 & &\!\!\!\! +{\bf j}(\alpha_1^{(0)} + \alpha_1^{(2)} +\alpha_1^{(4)} \cdots )
     + {\bf k}(\alpha_0^{(1)} + \alpha_0^{(3)} + \cdots ),
\end{array}
\label{GGFP-compo}
\end{equation}
where $\phi_1^{(0)}, \omega_0^{(1)}, B_1^{(2)}, \Omega_0^{(3)}, 
H_1^{(4)}, \cdots$ 
are bosonic gauge fields of 0-, 1-, 2-, 3-, 4-form, $\cdots$ with the form 
degree in the parentheses. Similarly $v_0^{(0)}, u_1^{(1)}, b_0^{(2)}, 
U_1^{(3)}, h_0^{(4)}, \cdots$ are bosonic gauge parameters of 
0-, 1-, 2-, 3-, 4-form, $\cdots$, respectively. Fermionic gauge fields 
and parameters are simply denoted by $\psi$ and $\alpha$ and the 
suffix ``$i$'' in $(i)$ denotes differential form degree. Hereafter we 
omit the differential form degree assignments for bosonic gauge fields 
and parameters for simplicity. 

Component expressions of the bosonic generalized Chern-Simons actions 
in 0-, 1-, 2-, 3-, 4-dimensions are given by 
\bea
S_0^k\!\!\!\!&=&\!\!\!\!\displaystyle
 \int \Tr\left[ -\phi_1(\hat{\psi}_0^{(0)})^2 - \frac{1}{3}\phi_1^3
\right], 
\nonumber 
\\
S_1^j\!\!\!\!&=&\!\!\!\!\displaystyle
\int\Tr\biggl[ 
\frac{1}{2}\phi_1(d \phi_1 + [\omega_0,\phi_1])
 - \hat{\psi}_0^{(0)}\{{\psi}_1^{(1)}, \phi_1 \} 
+\frac{1}{2}\hat{\psi}_0^{(0)}( d \hat{\psi}_0^{(0)}
+ [\omega_0,\hat{\psi}_0^{(0)}])\biggr],
\nonumber
\\
S_2^k \!\!\!\!&=&\!\!\!\!\displaystyle \int \Tr\biggl[
-\phi_1(d\omega_0 + \omega_0^2
+\{\hat{\psi}_0^{(0)},\hat{\psi}_0^{(2)}\} - ({\psi}_1^{(1)})^2)
\nonumber 
\\
&&
\hspace*{8.5mm}
-{\psi}_1^{(1)}(d \hat{\psi}_0^{(0)} +[\omega_0,\hat{\psi}_0^{(0)}])
- \phi_1^2 B_1 - (\hat{\psi}_0^{(0)})^2 B_1 \bigg],
\nonumber 
\\
S_3^j\!\!\!\!&=&\!\!\!\! \displaystyle
\int\Tr\bigg[-\frac{1}{2}\omega_0d\omega_0 -\frac{1}{3}\omega_0^3
+\hat{\psi}_0^{(0)}(d\hat{\psi}_0^{(2)} + [\omega_0,\hat{\psi}_0^{(2)}]
-\{{\psi}_1^{(1)},B_1\} -\{{\psi}_1^{(3)},\phi_1\})
\label{GCSA-compo}\\
&&
\hspace*{8.5mm}
+\phi_1(dB_1 + [\omega_0,B_1]) -\Omega_0(\phi_1^2
+(\hat{\psi}_0^{(0)})^2)\bigg],
\nonumber
\\
S_4^k \!\!\!\!&=&\!\!\!\!
\displaystyle\int\Tr\bigg[
-B_1(d\omega_0+\omega_0^2 + \{\hat{\psi}_0^{(0)},\hat{\psi}_0^{(2)}\}
-({\psi}_1^{(1)})^2)
- H_1((\hat{\psi}_0^{(0)})^2+\phi_1^2)
\nonumber
\\
&& 
\hspace*{8.5mm}
-\phi_1(d \Omega_0+\{\omega_0,\Omega_0\}+B_1^2 + 
\{\hat{\psi}_0^{(0)},\hat{\psi}_0^{(4)}\}
-\{{\psi}_1^{(1)},{\psi}_1^{(3)}\})
\nonumber
\\
&&
\hspace*{8.5mm}
- H_1((\hat{\psi}_0^{(0)})^2+\phi_1^2)
-{\psi}_1^{(1)}(d \hat{\psi}_0^{(2)}+ [\omega_0,\hat{\psi}_0^{(2)}]
+[\Omega_0,\hat{\psi}_0^{(0)}])-{\psi}_1^{(3)}[\omega_0,\hat{\psi}_0^{(0)}]
\bigg], 
\nonumber 
\eea
where $S^j_{(2k+1)}$ and $S^k_{(2k)}$ for $k=0,1,2, \cdots$ are, respectively, 
$(2k+1)$-dimensional and $(2k)$-dimensional actions.  
These generalized Chern-Simons actions are invariant under the following generalized 
gauge transformations for bosons:
\begin{equation}
\renewcommand{\arraystretch}{1.4}
\begin{array}{rcl}
\delta\phi_1\!\!\!\!&=&\!\!\!\!
[\phi_1,v_0] +  \{\hat{\psi}_0^{(0)},\hat{\alpha}_1^{(0)}\} ,
\\
\delta\omega_0\!\!\!\!&=&\!\!\!\! 
dv_0 + [\omega_0,v_0] +\{\phi_1,u_1\} +[\psi_1^{(1)},\hat{\alpha}_1^{(0)}]
-\{\hat{\psi}_0^{(0)},\alpha_0^{(1)}\} ,
\\
\delta B_1\!\!\!\!&=&\!\!\!\!
-du_1-\{\omega_0,u_1\}+[\phi_1,b_0]+[B_1,v_0] 
+\{\hat{\psi}_0^{(0)}, \hat{\alpha}_1^{(2)}\} + [\psi_1^{(1)},\alpha_0^{(1)}]
+\{\hat{\psi}_0^{(2)},\hat{\alpha}_1^{(0)}\} ,
\\
\delta\Omega_0\!\!\!\!&=&\!\!\!\!
db_0+[\omega_0,b_0] +\{\phi_1,U_1\} + \{B_1,u_1\} +[\Omega_0,v_0]
\\
&&\!\!\!\!
-\{\hat{\psi}_0^{(0)},\alpha_0^{(3)}\}+[\psi_1^{(1)},\hat{\alpha}_1^{(2)}]
-\{\hat{\psi}_0^{(2)},\alpha_0^{(1)}\}+[\psi_1^{(3)},\hat{\alpha}_1^{(0)}],
\\
\delta H_1\!\!\!\!&=&\!\!\!\! -dU_1 -\{\omega_0,U_1\} +[\phi_1,h_0] + [B_1,b_0] - \{
\Omega_0,u_1\}
+[H_1,v_0] 
\\
&&\!\!\!\!
+\{\hat{\psi}_0^{(0)},\alpha_1^{(4)}\}
+ [\psi_1^{(1)},\alpha_0^{(3)}] +\{\hat{\psi}_0^{(2)},\hat{\alpha}_1^{(2)}\}
+ [\psi_1^{(3)},\alpha_0^{(1)}] +\{\hat{\psi}_0^{(4)},\hat{\alpha}_1^{(0)}\},
\end{array}
\label{GGT-compo-boso}
\end{equation}
and for fermions:
\begin{equation}
\renewcommand{\arraystretch}{1.4}
\begin{array}{rcl}
\delta\hat{\psi}_0^{(0)}
\!\!\!\!&=&\!\!\!\! [\hat{\psi}_0^{(0)},v_0] - \{\phi_1,\hat{\alpha}_1^{(0)}\} ,
\\
\delta\psi_1^{(1)} 
\!\!\!\!&=&\!\!\!\! 
-d \hat{\alpha}_1^{(0)} - [\omega_0,\hat{\alpha}_1^{(0)}]
-[\phi_1,\alpha_0^{(1)}] - [\hat{\psi}_0^{(0)},u_1] + [\psi_1^{(1)},v_0] ,
\\
\delta\hat{\psi}_0^{(2)} 
\!\!\!\!&=&\!\!\!\!
d \alpha_0^{(1)} + \{\omega_0,\alpha_0^{(1)}\}
-\{\phi_1,\hat{\alpha}_1^{(2)}\} - \{B_1,\hat{\alpha}_1^{(0)}\} 
+[\hat{\psi}_0^{(0)},b_0] + [\psi_1^{(1)},u_1] +[\hat{\psi}_0^{(2)},v_0] ,
\\
\delta \psi_1^{(3)}
\!\!\!\!&=&\!\!\!\!
-d \hat{\alpha}_1^{(2)} -[\omega_0,\hat{\alpha}_1^{(2)}]
-[\phi_1,\alpha_0^{(3)}] - [B_1,\alpha_0^{(1)}] - [\Omega_0,\hat{\alpha}_1^{(0)}]
\\
&&\!\!\!\!
 -[\hat{\psi}_0^{(0)},U_1] +[\psi_1^{(1)},b_0] -[\hat{\psi}_0^{(2)},u_1] 
+[\psi_1^{(3)},v_0] ,
\\ 
\delta \hat{\psi}_0^{(4)}
\!\!\!\!&=&\!\!\!\!
d \alpha_0^{(3)} +\{\omega_0,\alpha_0^{(3)}\}
- \{\phi_1,\alpha_1^{(4)}\} - \{B_1,\hat{\alpha}_1^{(2)}\}
+ \{\Omega_0,\alpha_0^{(1)}\} 
-\{H_1,\hat{\alpha}_1^{(0)}\}
\\
&&\!\!\!\! +[\hat{\psi}_0^{(0)},h_0] + [\psi_1^{(1)},U_1] + [\hat{\psi}_0^{(2)},b_0]
+[\psi_1^{(3)},u_1] + [\hat{\psi}_0^{(4)},v_0].
\end{array}
\label{GGT-compo-fer}
\end{equation}

It is possible to derive component expressions of fermionic generalized 
Chern-Simons actions in 0-, 1-, 2-, 3-, 4-dimensions. We do not have an 
interpretation of the actions corresponding to $S^1$ and $S^i$ in (\ref{GCSAs}). 

\vspace*{3mm}

\noindent
{\bf 7 \  2-grading gauge system}

\vspace*{1mm}

\noindent
We have given a formulation of higher form gauge theory in terms of the 
3-grading structure with the quaternion algebra accommodating the 
generalized gauge system. We may wonder what happens if we consider only a 
2-grading structure. It turns out that the quaternion structure is kept to 
classify ${\mathcal A}$-type and ${\mathcal V}$-type generalized gauge 
fields and parameters. In the current assignment of the 3-grading structure, 
a 2-grading formulation can be derived by simply setting fermionic gauge 
fields and parameters to be zero: 
\beq
\psi = 0, \quad \alpha =0.  
\label{2-grading-sys}
\eeq

In this setting a bosonic version of generalized Chern-Simons actions can be
obtained with all the higher degrees of forms included. As one can see 
the leading terms of the actions in each dimensions have a typical 
$BF$-type structure except for 0 dimension. 

Let us specifically consider here the 3-, and 4-dimensional generalized 
Chern-Simons actions without fermions, which include the standard 
Chern-Simons action in 3 dimensions and $BF$ action in 4 dimensions, 
respectively: 
\begin{equation}
\renewcommand{\arraystretch}{2.0}
\begin{array}{rcl}
S_3^j(\psi=0)\!\!\!\!&=&\!\!\!\!
\displaystyle \int\Tr\bigg[-\frac{1}{2}\omega_0d\omega_0
-\frac{1}{3}\omega_0^3
+\phi_1(dB_1 + [\omega_0,B_1]) -\Omega_0\phi_1^2 \bigg],
\\
S_4^k(\psi=0) \!\!\!\!&=&\!\!\!\!
\displaystyle\int\Tr\bigg[
-B_1(d\omega_0+\omega_0^2 )
-\phi_1(d \Omega_0+\{\omega_0,\Omega_0\}+B_1^2) - H_1 \phi_1^2 \bigg],
\end{array}
\label{3-4-GCSAs}
\end{equation}
which are invariant under the following generalized gauge transformations:
\begin{equation}
\renewcommand{\arraystretch}{1.4}
\begin{array}{rcl}
\delta\phi_1\!\!\!\!&=&\!\!\!\! [\phi_1,v_0],
\\
\delta\omega_0\!\!\!\!&=&\!\!\!\! 
dv_0 + [\omega_0,v_0] +\{\phi_1,u_1\},
\\
\delta B_1\!\!\!\!&=&\!\!\!\!
-du_1-\{\omega_0,u_1\}+[\phi_1,b_0]+[B_1,v_0],
\\
\delta\Omega_0\!\!\!\!&=&\!\!\!\!
db_0+[\omega_0,b_0] +\{\phi_1,U_1\} + \{B_1,u_1\}
+[\Omega_0,v_0], \\
\delta H_1\!\!\!\!&=&\!\!\!\!
-dU_1 -\{\omega_0,U_1\} +[\phi_1,h_0] + [B_1,b_0]
- \{\Omega_0,u_1\} +[H_1,v_0]. 
\end{array}
\label{GGT-compo-no-ferm}
\end{equation}
One can easily see that these generalized actions are natural 
generalizations of the standard Chern-Simons action in 3 and 4 dimensions 
that include all the degrees of differential form. 
The 4-dimensional generalized Chern-Simons action includes the $BF$ action 
as a leading term. We can interpret this formulation as a 
generalization of the Chern-Simons action to arbitrary dimensions 
and thus we call these actions generalized Chern-Simons actions. 
It should be noted here that in the generalized gauge transformations 
(\ref{GGT-compo-no-ferm}) commutators and anti-commutators are mixed. 
However after taking into account the odd-form nature and the odd-grading 
nature of the suffix ``1'', all the anti-commutators turn into commutators 
so that algebra is closed within the Lie algebra as it was already noted after 
Eqs. (\ref{GGTs}).  

It is important to realize that all the even-form generalized gauge fields 
in (\ref{3-4-GCSAs}) and all the odd-form generalized gauge parameters 
in (\ref{GGT-compo-no-ferm}) carry the suffix $1$. 
The even-form gauge fields and the odd-form gauge parameters have a 
hidden grading structure, whose nature is not completely determined and 
could also be interpreted as a grading of fermionic nature.

\vspace*{3mm}

\noindent
{\bf 8 \ Equations of motions for generalized Chern-Simons actions}

\vspace*{1mm}

\noindent
The equations of motion for the generalized Chern-Simons actions can be 
derived and are given by the vanishing condition for the generalized curvature:
$$
{\mathcal F} = Q{\mathcal A}+{\mathcal A}^2
={\bf 1}{\mathcal F}^1+{\bf i}{\mathcal F}^i+{\bf j}{\mathcal F}^j 
+ {\bf k}{\mathcal F}^k = 0,
$$
with 
\begin{equation}
\renewcommand{\arraystretch}{1.4}
\begin{array}{rcl}
{\mathcal F}^1\!\!\!\!&=&\!\!\!\!
-dA_0 - A_0^2 - \hat{A}_1^2+ \psi_1^2 - \hat{\psi}_0^2 = 0,
\\
{\mathcal F}^i \!\!\!\!&=&\!\!\!\!
d\hat{A}_1 + [A_0, \hat{A}_1] + \{\psi_1,\hat{\psi}_0\}= 0,
\\
{\mathcal F}^j \!\!\!\!&=&\!\!\!\!
d\psi_1 + \{A_0, \psi_1 \} + [\hat{A}_1, \hat{\psi}_0] = 0,
\\
{\mathcal F}^k \!\!\!\!&=&\!\!\!\! -d\hat{\psi}_0 -[A_0,\hat{\psi}_0] +
 \{\hat{A}_1,\psi_1\}=0.
\end{array}
\label{GEMs}
\end{equation}
Notice that the first line of (\ref{GEMs}) is a generalization of the 
Maurer-Cartan equation and it admits the solution 
\beq
{\mathcal A} = {\mathcal G}^{-1} Q {\mathcal G},  
\eeq
with 
\beq
{\mathcal G} = e^{i{\mathcal V}},
\eeq
where ${\mathcal V}$ is a generalized parameter as in 
Eq. (\ref{gene-nonAbelian}).

Component expressions for the equations of motion can be derived by 
equating to zero separately the sectors of the generalized curvature 
with different form degree. 
Bosonic equations of motion for 0-, 1-, 2-, 3-, 4-form sectors are given by:
\begin{equation}
\renewcommand{\arraystretch}{1.4}
\begin{array}{rcl}
-\phi_1^2 -(\hat{\psi}_0^{(0)})^2 \!\!\!\!&=&\!\!\!\!
0, 
\\
d\phi_1 +[\omega_0,\phi_1] +\{\psi_1^{(1)},\hat{\psi}_0^{(0)}\}
\!\!\!\!&=&\!\!\!\!
0,
\\
-d\omega_0 -\omega_0^2 - \{\phi_1,B_1\} + (\psi_1^{(1)})^2
-\{\hat{\psi}_0^{(0)},\hat{\psi}_0^{(2)}\} \!\!\!\!&=&\!\!\!\! 
0,
\\
dB_1 + [\omega_0,B_1] + [\Omega_0,\phi_1] 
+ \{\psi_1^{(1)},\hat{\psi}_0^{(2)}\} + \{\psi_1^{(3)},\hat{\psi}_0^{(0)}\}
\!\!\!\!&=&\!\!\!\! 
0, 
\\
-d\Omega_0 -\{\omega_0,\Omega_0\} -\{\phi_1,H_1\} - B_1^2
+ \{\psi_1^{(1)},\psi_1^{(3)}\} - \{\hat{\psi}_0^{(0)},\hat{\psi}_0^{(4)}\}
-(\hat{\psi}_0^{(2)})^2 \!\!\!\!&=&\!\!\!\! 0.
\end{array}
\label{GEM-compo-boson}
\end{equation}
And corresponding fermionic equations of motion are
\begin{equation}
\renewcommand{\arraystretch}{1.4}
\begin{array}{rcl}
[\phi_1, \hat{\psi}_0^{(0)}] \!\!\!\!&=&\!\!\!\!0,
\\
-d\hat{\psi}_0^{(0)} -[\omega_0,\hat{\psi}_0^{(0)}] + \{\phi_1,\psi_1^{(1)}\}
\!\!\!\!&=&\!\!\!\!0,
\\
d\psi_1^{(1)} + \{\omega_0,\psi_1^{(1)}\} +[\phi_1,\hat{\psi}_0^{(2)}]
+ [B_1,\hat{\psi}_0^{(0)}] \!\!\!\!&=&\!\!\!\! 0,
\\
-d\hat{\psi}_0^{(2)} - [\omega_0,\hat{\psi}_0^{(2)}] +\{\phi_1,\psi_1^{(3)}\}
+\{B_1,\psi_1^{(1)}\} -[\Omega_0,\hat{\psi}_0^{(0)}]
\!\!\!\!&=&\!\!\!\! 0,
\\
d\psi_1^{(3)} +\{\omega_0,\psi_1^{(3)}\} + [\phi_1,\hat{\psi}_0^{(4)}]
+[B_1,\hat{\psi}_0^{(2)}] +\{\Omega_0,\psi_1^{(1)}\} +[H_1,\hat{\psi}_0^{(0)}]
\!\!\!\!&=&\!\!\!\! 0.
\end{array}
\label{GEM-compo-ferm}
\end{equation}
The equations of motion of 3- and 4-dimensional generalized Chern-Simons 
actions in (\ref{3-4-GCSAs}) are simply given by setting $\psi=0$ in 
(\ref{GEM-compo-boson}). It is interesting to realize that these generalized 
equations of motion have the generalized gauge invariance. 

\vspace*{3mm}

\noindent
{\bf 9 \ Generalized topological relations}

\vspace*{1mm}

\noindent
A generalized gauge theory version of the second Chern character can be 
defined and it is related to the generalized Chern-Simons action 
in analogy with the standard gauge theory: 
\beq
\int \Tr({\mathcal F}^2) = \int \Tr Q\left( {\mathcal A}Q{\mathcal A} 
+\frac{2}{3}{\mathcal A}^3 \right) \equiv
{\bf 1}{\mathcal G}^1 + {\bf i}{\mathcal G}^i + {\bf j}{\mathcal G}^j 
+ {\bf k}{\mathcal G}^k ,
\label{top-rel} 
\eeq
where ${\mathcal F}$, ${\mathcal A}$ and $Q$ are defined (\ref{GEMs}), 
(\ref{GGFP-compo}) and (\ref{gene-d-op}), respectively. 

From the sector-wise equivalence for the quaternion in (\ref{top-rel}) 
we can obtain the following new topological relations: 
\begin{equation}
\renewcommand{\arraystretch}{2.0}
\begin{array}{rcl}
{\mathcal G}^1 \!\!\!\!&\equiv&\!\!\!\!
\displaystyle\int \Tr\left(({\mathcal F}^1)^2 - ({\mathcal F}^i)^2 
- ({\mathcal F}^j)^2 - ({\mathcal F}^k)^2\right) =
-2\int\Tr\,  d{\mathcal L}_{GCS}^j, 
\\
{\mathcal G}^i \!\!\!\!&\equiv&\!\!\!\!
\displaystyle\int\Tr\left( \{{\mathcal F}^1, {\mathcal F}^i\}
+[{\mathcal F}^j,{\mathcal F}^k] \right) =
2\int\Tr\, d{\mathcal L}_{GCS}^k,
\\
{\mathcal G}^j \!\!\!\!&\equiv&\!\!\!\!
\displaystyle\int\Tr\left( \{{\mathcal F}^1, {\mathcal F}^j\}
+[{\mathcal F}^k,{\mathcal F}^i] \right) =
2\int\Tr\, d{\mathcal L}_{GCS}^1,
\\
{\mathcal G}^k \!\!\!\!&\equiv&\!\!\!\!
\displaystyle
\int\Tr\left( \{{\mathcal F}^1, {\mathcal F}^k\}
+[{\mathcal F}^i,{\mathcal F}^j] \right) =
-2\int\Tr\, d{\mathcal L}_{GCS}^i,
\end{array}
\label{top-rel-compo}
\end{equation}
where ${\mathcal F}^A\,  (A=1,i,j,k)$ are given in (\ref{GEMs}) 
and ${\mathcal L}_{GCS}^A \, (A=1,i,j,k)$ are given by the following relations:
\beq
S^A =\int \Tr {\mathcal L}_{GCS}^A, \qquad (A=1,i,j,k), 
\eeq
with $S^A$ given in (\ref{GCSAs}). 

It should be recognized here that $\Tr({\mathcal F}^2)$ and correspondingly 
the right hand side in (\ref{top-rel}) have the ${\mathcal V}$-type 
quaternion structure. Therefore ${\mathcal G}^1$ and 
${\mathcal G}^i$ are bosonic even- and odd-form sectors while ${\mathcal G}^j$ 
and ${\mathcal G}^k$ are fermionic even- and odd-form sectors respectively. 

Reduction to 2-grading formulation which includes all degrees of differential 
forms for bosonic gauge fields with hidden grading for even-forms, 
can be easily obtained by simply setting $\psi=0$ in ${\mathcal G}^1$ 
and ${\mathcal G}^i$ while ${\mathcal G}^j$ and ${\mathcal G}^k$ disappear. 
For a given space-time dimension the expressions of ${\mathcal G}^A$ can 
be derived by extracting in (\ref{top-rel-compo}) the sector of the 
corresponding from degree.
For example 4-dimensional counter part of the generalized topological relation 
without fermions is given by 
\bea
{\mathcal G}^1_4 \!\!\!\!&=&\!\!\!\! 
\int\Tr\bigg[(d\omega_0 + \omega_0^2 + 
\{\phi_1,B_1\})^2 
+ \{\phi_1^2,~ d\Omega_0+\{\omega_0,\Omega_0\}+B_1^2+\{\phi_1,H_1\}\}
\nonumber \\
&&\hspace*{10mm}
-\{d\phi_1+[\omega_0,\phi_1],~dB_1+[\omega_0,B_1]+[\Omega_0,\phi_1]\} 
\bigg] 
\nonumber\\
\!\!\!\!&=&\!\!\!\!
\int\Tr\, d\bigg[\omega_0d\omega_0 + \frac{2}{3}\omega_0^3
-2\phi_1(dB_1+[\omega_0,B_1]) + 2\Omega_0\phi_1^2
\bigg], 
\label{topo-rel-compo}
\eea
where the last term is the exterior derivative times the 3-dimensional 
generalized Chern-Simons action in (\ref{3-4-GCSAs}). It should be noted that 
0-, 2-, 4-form gauge fields $\phi_1,~B_1$ and $H_1$ carry a hidden 
$\mathbb{Z}_2$-grading. Notice also that some nontrivial cancellations occur in going 
from the second to the third term due to the hidden grading. 
It is interesting to realize that the topological Yang-Mills action is 
included as the first term in the action in (\ref{topo-rel-compo}).

\vspace*{3mm}

\noindent
{\bf 10 \ Generalized differential operators}

\vspace*{1mm}

\noindent
In the component expressions for the 
actions given in (\ref{GCSA-compo}) we have specifically assigned the first 
and the second grading as 
(fermion, boson) and (even form, odd form) so that the second grading is 
related to the space-time structure of the generalized fields and parameters. 
There is some freedom in the interpretation of gradings. 
A physical meaning of the third grading is unspecified in this paper. 
This assignment of gradings automatically generates an asymmetric nature 
of exterior derivative operator with respect to the quaternion structure 
as in (\ref{gene-d-op}). This special assignment is the origin of the 
difference between the action $S^j$ and the rest of the actions and 
the similarity of $S^1,S^i$ and $S^k$, as is already pointed out. 

We have introduced fermionic fields which have an anti-symmetric 
space-time tensor structure. There is no spinor introduced. 
One may wonder what kind of fermions they are. It was shown that 
these fermions can be identified as ghost fields of quantized 
generalized Chern-Simons, which turned out to 
be an infinitely reducible gauge system~\cite{GCS-quant}. 
It was noticed that the generalized differential operator can be 
extended to:
\beq
Q=-{\bf i}s + {\bf j}d, 
\label{extended-Q1}
\eeq
where $s$ is a fermionic zero form and can be interpreted as a 
BRST operator. 

The generalized Chern-Simons action (\ref{GCSA}) has the gauge 
invariance with the generalized gauge parameters which turn into 
ghosts after quantization. It is surprising to find that 
${\mathcal V}$-type gauge parameters turn into ghosts as 
${\mathcal A}$-type gauge fields due to the change into 
alternative grading for fermion and boson. 
This structure is realized as a BRST 
transformation by the addition of $-{\bf i}s$ term for the 
$Q$-operator in (\ref{extended-Q1}). It is fundamentally reflected 
from the fact that the generalized gauge transformation is 
infinitely reducible~\cite{GCS-quant}.

The quantization of the generalized Chern-Simons action\cite{GCS-quant} 
was completed by BV and BFV formalism~\cite{BV,BFV}. BV formalism is 
the special realization of Q-P manifold, {\it \`a la} 
AKSZ formalism~\cite{AKSZ}. The connection of the current formulation 
of the generalized gauge theory and the AKSZ formalism is of a great 
interest. It is expected that the topological particle field 
theory formulation which was discovered to be equivalent to the 
quaternion formulation of the generalized gauge theory plays 
an important role~\cite{TPFT}.

As a generalization of the $Q$ operator (\ref{extended-Q1}) 
we propose the most symmetric differential operator: 
\beq
Q=-{\bf 1}sd\sigma -{\bf i}s + {\bf j}d +{\bf k}\sigma,
\label{extended-Q2}
\eeq
where $\sigma$ is an exterior derivative operator corresponding 
to the 3rd grading. This $Q$ operator satisfies the item 1. and 2. 
in section 4 and thus define a more general and symmetric generalized 
gauge theory. 


\vspace*{3mm}

\noindent
{\bf 11 \  Conclusions and discussions}

\vspace*{1mm}

\noindent
The generalized Chern-Simons gauge theory to accommodate all the degrees 
of differential forms as Lie algebra valued gauge fields and parameters 
is explicitly formulated for arbitrary even and odd dimensions. 
The role of quaternion in the formulation is clarified as an alternative 
presentation for three grading structure of the generalized gauge fields 
and parameters. The connection of the Lie algebra formulation with the 
original graded Lie algebra formulation is also clarified. 

In considering a gauge system of three physically independent gradings, 
it is natural to assume that these gradings commute each other. 
For example the differential form grading of space-time and the 
quantized ghost grading have no physical connection in general. 
Thus we assume that these gradings are commutative as in 
(\ref{grading1}). In this physical 
system the quaternion accompanied formulation of the generalized 
gauge theory is necessary. On the other hand this formulation can 
be equivalently described by the formulation of the total three 
gradings of (\ref{grading2}). 

There has been recently a growing interest in higher form gauge 
theory as a way of finding a unified formulation of extended objects 
such as D-branes~\cite{string-related1,string-related2,string-related3} 
and of getting a deeper understanding of duality 
related prescriptions for gauge theories~\cite{Seiberg}. 
We consider that the present formulation as well as its original 
version~\cite{kawa-wata} have possible applications in 
this new context and also in the older formulations of topological field 
theories~\cite{BF}, Chern-Simons related 
gravities~\cite{Witten-3dgravity,kawa-wata2,CSgf-ext}, and 
higher spin formulations~\cite{higher-spin}. 
It is also known that the standard Chern-Simons and BF actions 
lead simplicial gravities of Ponzano-Regge type in 3 dimensions 
and 15-$j$ topological gravity type in 4 
dimensions~\cite{PR-G,BF-G}.
One may wonder what is the role of the extra differential forms 
other than the ones constructing gravity background. 
The generalized gauge theory formulation suggests a hope 
that we may be able to formulate a simplicial gravity theory 
with matter introduced naturally on the simplicial lattice 
by simplex-form correspondence~\cite{NK-R}. 

Finally we would like to point out that physical identification 
of the unknown third grading would be very important for 
finding the generalized gauge symmetry.




\end{document}